\documentclass[10pt,twoside,a4paper]{article}
\usepackage{fullpage}
\usepackage[affil-it]{authblk}

\usepackage{natbib,aas_macros}
\usepackage{graphicx}
\usepackage{amssymb}
\usepackage{psfrag}

\title{Modelling the 21~cm Signal From the Epoch of Reionization and Cosmic Dawn}
\author[1]{T. Roy Choudhury\thanks{\tt tirth@ncra.tifr.res.in}}
\author[2]{Kanan Datta}
\author[3]{Suman Majumdar}
\author[1]{Raghunath Ghara}
\author[4]{Aseem Paranjape}
\author[5]{Rajesh Mondal}
\author[5]{Somnath Bharadwaj}
\author[2]{Saumyadip Samui}

\affil[1]{National Centre for Radio Astrophysics, TIFR, Post Bag 3, Ganeshkhind, Pune 411007, India}
\affil[2]{Department of Physics, Presidency University, 86/1 College Street, Kolkata - 700073, India}
\affil[3]{Department of Physics, Blackett Laboratory, Imperial College, London SW7 2AZ, UK}
\affil[4]{Inter-University Center for Astronomy \& Astrophysics, Post Bag 4, Ganeshkhind, Pune 411007, India}
\affil[5]{Department of  Physics \& Centre for Theoretical Studies, Indian Institute of Technology Kharagpur, Kharagpur - 721302, India}

\begin{document}

\date{}

\maketitle

\thispagestyle{empty}

\begin{abstract}
Studying the cosmic dawn and the epoch of reionization through the redshifted 21~cm line are among the major science goals of the SKA1. Their significance lies in the fact that they are closely related to the very first stars in the universe. Interpreting the upcoming data would require detailed modelling of the relevant physical processes. In this article, we focus on the theoretical models of reionization that have been worked out by various groups working in India with the upcoming SKA in mind. These models include purely analytical and semi-numerical calculations as well as fully numerical radiative transfer simulations. The predictions of the 21~cm signal from these models would be useful in constraining the properties of the early galaxies using the SKA data.


\end{abstract}

\section{Introduction}

One of the major science goals of the SKA is to study the redshifted 21 cm signal of neutral hydrogen from the cosmic dawn and the epoch of reionization \citep{2013ExA....36..235M,2015aska.confE...1K,2015aska.confE.171C}. The cosmic dawn refers to a period when the first stars formed in the universe, while the epoch of reionization is when the HI in the intergalactic medium was being ionized by the UV radiation from the first stars \citep[see, e.g.,][]{2001PhR...349..125B}. It is believed that this is process which extended over redshift ranges $15 \gtrsim z \gtrsim 6$ \citep{2015MNRAS.454L..76M}. The redshifted 21 cm radiation is possibly the most promising method of detecting the distribution of the HI at these redshifts \citep{2006PhR...433..181F,2012RPPh...75h6901P}. The signal will contain information about how this process occurred, and will indirectly help in studying the properties of the first stars and the surrounding medium at those early epochs.

Detailed models of reionization are a crucial ingredient for interpreting the data and constraining the EoR. The difficulty in constructing models at such high redshifts is that there are practically no observational constraints on the physics of galaxy formation. As a result it is often not possible to have a good idea about the number of ionizing photons available. The models thus assume simple prescriptions to assign ionizing luminosities to dark matter haloes, and try to constrain the free parameters using existing observations.

This article is meant to highlight the advances made in modelling the EoR and cosmic dawn by members of the Indian community keeping the upcoming SKA in mind. It turns out that groups working in related areas have contributed to different types of modelling, starting from purely analytical ones to complex radiative transfer simulations. The main features of these models and their main results are summarised in the following sections.

\section{Present constraints on reionization history}

The main components of building a reionization model are as follows:
\begin{itemize}

\item The abundance (and locations) of dark matter haloes form the first step as the sources of ionizing photons (galaxies) form within these haloes. In analytical models, one can use the standard forms of the halo mass function $d n(M,z) / d M$ given by, e.g., \citet{1974ApJ...187..425P,1991ApJ...379..440B,1999MNRAS.308..119S}, while in the $N$-body simulations, the masses and locations of these haloes are usually obtained by group finder algorithms, e.g., the Friends-of-friend \citep{1985ApJ...292..371D}.

\item Given the dark matter haloes, one needs to work out the physical processes related to galaxy formation, stellar radiation and escape of ionizing photons. All these are highly uncertain at $z \gtrsim 6$, hence most reionization models tend to assume some prescription which relates the number of ionizing photons to the halo mass. The simplest of these assume the number of ionizing photons in the IGM produced by a halo of mass $M$ to be given by
\begin{equation}
N_{\gamma} = \zeta \left(\frac{\Omega_H}{\Omega_m}\right)~\frac{M}{m_H},
\end{equation}
where $\zeta$ is an unknown parameter and all other symbols have their usual meanings. Physically $\zeta$ is a combination of star-forming efficiency, the number of ionizing photons produced by stars and the fraction of ionizing photons that escape from the host galaxy into the IGM. The above relation can also be written in terms of the number of photons in the IGM per unit time per unit comoving volume
\begin{equation}
\dot{n}_{\gamma} = \zeta n_H \frac{d f_{\rm coll}}{d t},
\end{equation}
where $n_H$ is the comoving number density of hydrogen and $f_{\rm coll}$ is the mass fraction in collapsed haloes (called the collapsed fraction) that are forming stars. The collapsed fraction is related to the mass function by
\begin{equation}
f_{\rm coll} = \frac{1}{\bar{\rho}_m} \int_{M_{\rm min}}^{\infty} d  M ~ M~ \frac{d n(M,z)}{d M},
\end{equation}
where $M_{\rm min}$ is the minimum mass of haloes that can form stars. The value of $M_{\rm min}$ is decided by the cooling efficiency of the gas in collapsed haloes, e.g., if the gas contains only atomic hydrogen, it is unable to cool at virial temperatures lower than $10^4$ K while the presence of molecules can push the limit to $\sim 300$ K.

\item The third component is a description of the inhomogeneous baryonic density field. In analytical models, this could be described by the PDF $P(\Delta_B)$ of the baryonic overdensity $\Delta_B$ (smoothed over the Jeans scale). Some standard forms that are used are the lognormal distribution \citep{1997ApJ...479..523B,2001MNRAS.322..561C,2001ApJ...559...29C,2005MNRAS.361..577C}, or a fitting form motivated by the hydrodynamical simulations \citep{2000ApJ...530....1M,2009MNRAS.398L..26B}. While simulating the baryonic densities, it is important to be able to resolve reasonably small scales so as to identify the dense optically thick systems. These regions, because of their high recombination rate, act as ``sinks'' of ionizing photons and can alter the distribution of ionized regions.

\item Finally, given the ionization sources and the clumpy baryonic field, one has to solve the transfer of radiation through the IGM accounting for all relevant physical processes. The cosmological radiative transfer equation
\begin{equation}
\frac{\partial I_{\nu}}{\partial t} + \frac{c}{a(t)} {\bf \hat{n} \cdot \nabla_x} I_{\nu}
-H(t) \nu \frac{\partial I_{\nu}}{\partial \nu} 
+ 3 H(t) I_{\nu} = -c \kappa_{\nu} I_{\nu} + \frac{c}{4 \pi} \epsilon_{\nu},
\label{eq:radtrans_local}
\end{equation}
where $I_{\nu} \equiv I(t, {\bf x}, {\bf n}, \nu)$ is the monochromatic specific intensity of the radiation field, ${\bf n}$ is a unit vector along the direction of propagation of the radaiation, $\kappa_{\nu}$ is the absorption coefficient and $\epsilon_{\nu}$ is the emissivity, has to be solved at every point in the seven-dimensional $(t,{\bf x}, {\bf n},\nu)$ space. However, the high dimensionality of the problem makes the solution of the complete radiative transfer equation well beyond our capabilities, particularly since we do not have any obvious symmetries in the problem and often need high spatial and angular resolution in cosmological simulations. Hence, the approach to the problem has been to use different numerical schemes and approximations \citep{2006MNRAS.371.1057I}.

In analytical models, one can simplify the problem by taking the global average of equation (\ref{eq:radtrans_local}) under the assumption that the mean free path of photons is significantly smaller than the horizon size \citep{2009CSci...97..841C}. The equation can be written in terms of the volume filling factor $Q_{\rm HII}$ of ionized regions as
\begin{equation}
\frac{d Q_{\rm HII}}{d t} = \frac{\dot{n}_{\gamma}}{n_H} - Q_{\rm HII} ~ {\cal C}~ \alpha(T)~ n_H,
\label{eq:dQdt}
\end{equation}
where ${\cal C}$ is the clumping factor and $\alpha(T)$ is the recombination coefficient at a temperature $T$.

\end{itemize}

\begin{figure}
\begin{center}
\includegraphics[width=0.8\textwidth]{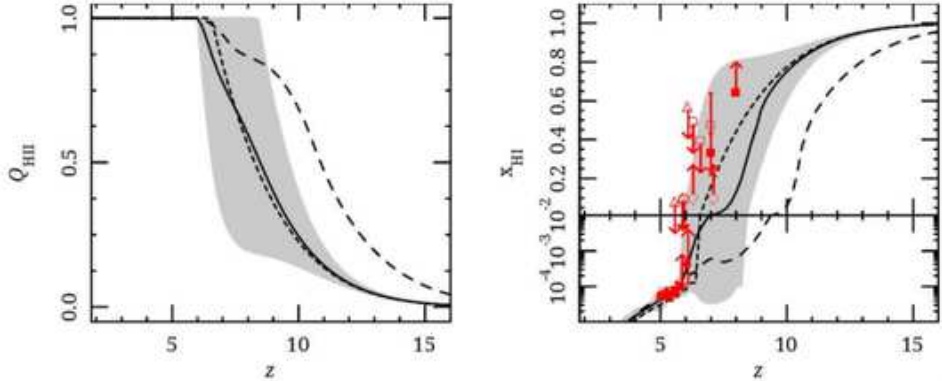}
\end{center}
\caption{The mean value (solid lines) and its 2$\sigma$ limits (shaded regions) for the ionized volume fraction (left panel) and the neutral hydrogen fraction (right panel) obtained using a PCA + MCMC analysis with Planck data \cite{}. The fiducial model (short-dashed lines) and the model constrained using WMAP9 data (long-dashed lines) are also shown for comparison. We also show the observational limits on neutral hydrogen fraction $x_{\rm H I} (z)$ (right panel) from various measurements, for details see \citet{2015MNRAS.454L..76M}.}
\label{fig:PCA_constraints}
\end{figure}

The uncertainties in the reionization models can, in principle, be constrained by comparing with existing data, in particular, the quasar absorption spectra blueward of the Ly$\alpha$ emission line and the measurements of the Thomson scattering optical depth from the CMB observations. One such physically motivated semi-analytical model \citep{2011MNRAS.413.1569M,2012MNRAS.419.1480M,2015MNRAS.454L..76M} which treats the $\zeta$ as an unknown function of $z$ and constraints it using a Principal Component Analysis, predicts the evolution of $Q_{\rm HII}$ and the average neutral fraction $x_{\rm HI}$ as shown in Figure \ref{fig:PCA_constraints}. One can see that the present data sets imply that reionization begins at $z \sim 15$ and is completed close to $z \sim 6$. There is still considerable uncertainty around the final stages of reionization as is shown by the width of the shaded regions around $z \sim 7-8$, and one expects the 21~cm experiments to play an important role in reducing the uncertainties.

\begin{figure}
\centering
\includegraphics[width=0.55\textwidth]{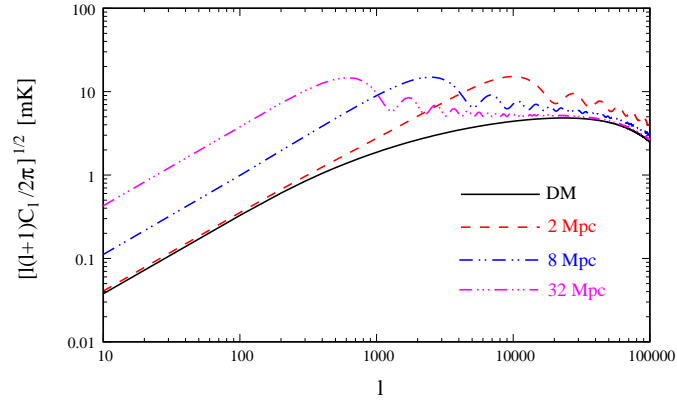}
\caption{The angular power spectrum of HI brightness temperature fluctuations for a model with non-overlapping spherical bubbles of a fixed radius. The results are shown for three different values of the radius. Also shown is the power spectrum for dark matter linear fluctuations. See \citet{2007MNRAS.378..119D} for details.}
\label{fig:cl_n}
\end{figure}

\begin{figure*}
\centering
\includegraphics[angle=270,width=0.6\textwidth]{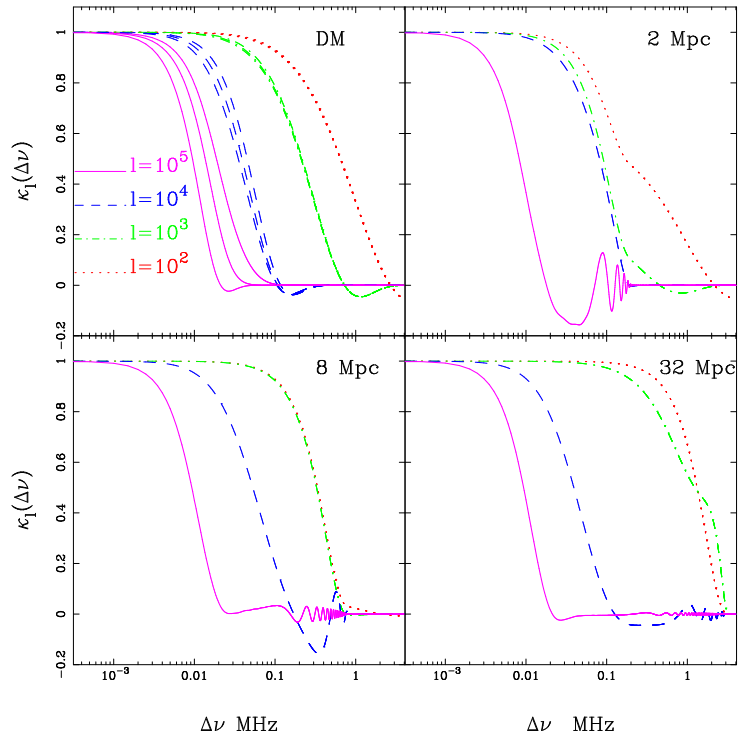}
\caption{The frequency decorrelation function for four values of $l$. Results are shown for the dark matter fluctuations (upper left panel) and the reionization model with non-overlapping spherical bubbles of a fixed radius (the other three panels). See \citet{2007MNRAS.378..119D} for details.}
\label{fig:kappa_l}
\end{figure*}

\section{Modelling the 21~cm signal from EoR}

The basic principle which is central to the 21~cm experiments is the neutral hydrogen hyperfine transition line at a rest wavelength of 21~cm. This line, when redshifted, is observable at low frequencies ($\sim 90 - 200$ MHz for $z \sim 15 - 6$) against the CMB. The strength of the signal is quantified by the differential brightness temperature given by
\begin{equation}
\delta T_b = \bar{T}~x_{\rm HI}~\Delta_B~\left(\frac{T_S - T_{\rm CMB}}{T_S}\right),
\label{eq:delta_T_b}
\end{equation}
where $T_S$ is the spin temperature of the gas and
\begin{equation}
\bar{T} = 27 \mbox{ mK} \left(\frac{\Omega_B h^2}{0.023}\right) \left(\frac{0.15}{\Omega_m} \frac{1+z}{10}\right)^{1/2}.
\end{equation}
The quantities $x_{\rm HI}$ and $\Delta_B$ are the neutral hydrogen fraction and the baryonic overdensity respectively. The radio-interferometric observations would measure only the fluctuations in $\delta T_b$, e.g., the power spectrum $P(\mathbf{k})$ defined as
\begin{equation}
\langle \delta \hat{T}_b(\mathbf{k})~\delta \hat{T}^*_b(\mathbf{k'})\rangle = (2 \pi)^3 \delta_D(\mathbf{k - k'}) P(\mathbf{k}),
\end{equation}
where $\delta \hat{T}_b(\mathbf{k})$ is Fourier transform of $\delta T_b$. As can be seen from equation (\ref{eq:delta_T_b}), the fluctuations in the signal are sourced by either fluctuations in the neutral hydrogen density field $x_{\rm HI}~\Delta_B$ or in the spin temperature $T_S$. Unless one is interested in the very early stages of reionization (cosmic dawn), the IGM can be taken to be sufficiently heated by X-rays and the $T_S$ coupled to the gas temperature through the Ly$\alpha$ coupling \citep{1952AJ.....57R..31W,1959ApJ...129..536F}. In that case $T_S \gg T_{\rm CMB}$ and hence the signal is simply $\delta T_b \propto x_{\rm HI}~\Delta_B$, i.e., the signal will simply follow the neutral hydrogen distribution.

The above expressions do not account for the line of sight peculiar velocities of the gas which can make the power spectrum anisotropic. Other sources of anisotropies are the light cone effect and the Alcock–Paczynski effect. In that case, the power spectrum can be denoted as $P(k, \mu)$, where $\mu$ is the direction cosine between the line of sight and $\mathbf{k}$. It is expected that the first generation of 21~cm experiments would measure the spherically averaged power spectrum $P_0(k) = \int_{-1}^1 (d \mu/2)~P(k,\mu)$.

A related quantity of interest which can be measured from the observation is the multi-frequency angular power spectrum (MAPS) defined as \citep{2007MNRAS.378..119D}
\begin{equation}
C_l(\Delta \nu) = \frac{1}{\pi r_{\nu}^2} \int_0^{\infty} d k_{\parallel}~\cos(k_{\parallel} r'_{\nu} \Delta \nu)~P(\mathbf{k}),
\end{equation}
where $r_{\nu}$ is the comoving distance to the redshift $z = 1420 \mbox{ MHz} / \nu - 1$ and $\mathbf{k}$ has a magnitude $k = \sqrt{k_{\parallel}^2 + l^2 / r_{\nu}^2}$. The quantity $C_l(0)$ is essentially the two-dimensional power spectrum on a plane at the distance $r_{\nu}$ from the observer.

\subsection{Analytical models}

Analytical models of reionization are based on modelling the size distribution of ionized bubbles around galaxies, which then can be extended to obtain the power spectrum. The simplest model would be to approximate the ionization field as a collection of non-overlapping sphere of fixed radius $R$ \citep{2005MNRAS.356.1519B,2007MNRAS.378..119D}. The values of $R$ and the number density of such bubbles $n_{\rm bub}$ can be chosen to obtain a particular ionized fraction $Q_{\rm HII}(z) = n_{\rm bub}~4 \pi R^3 / 3$ at a given epoch. The quantity $C_l(0)$ for different values of $R$ for such a model is shown in Fig \ref{fig:cl_n}. One can see that the signal is enhanced compared to the underlying dark matter fluctuations at angular scales larger than the bubble size. It peaks around a value of $l$ which is inversely proportional to the angular size of the bubbles. The frequency decorrelation function $\kappa_l(\Delta \nu) = C_l(\Delta \nu) / C_l(0)$ as function of $\Delta \nu$ is plotted in Fig \ref{fig:kappa_l} which shows that the fluctuations decorrelate quite rapidly particularly for large $l$. This decorrelation can be useful in separating the cosmological signal from other astrophysical foregrounds which have relatively smooth frequency spectra.

\begin{figure}
\centering
\includegraphics[width=0.9\textwidth]{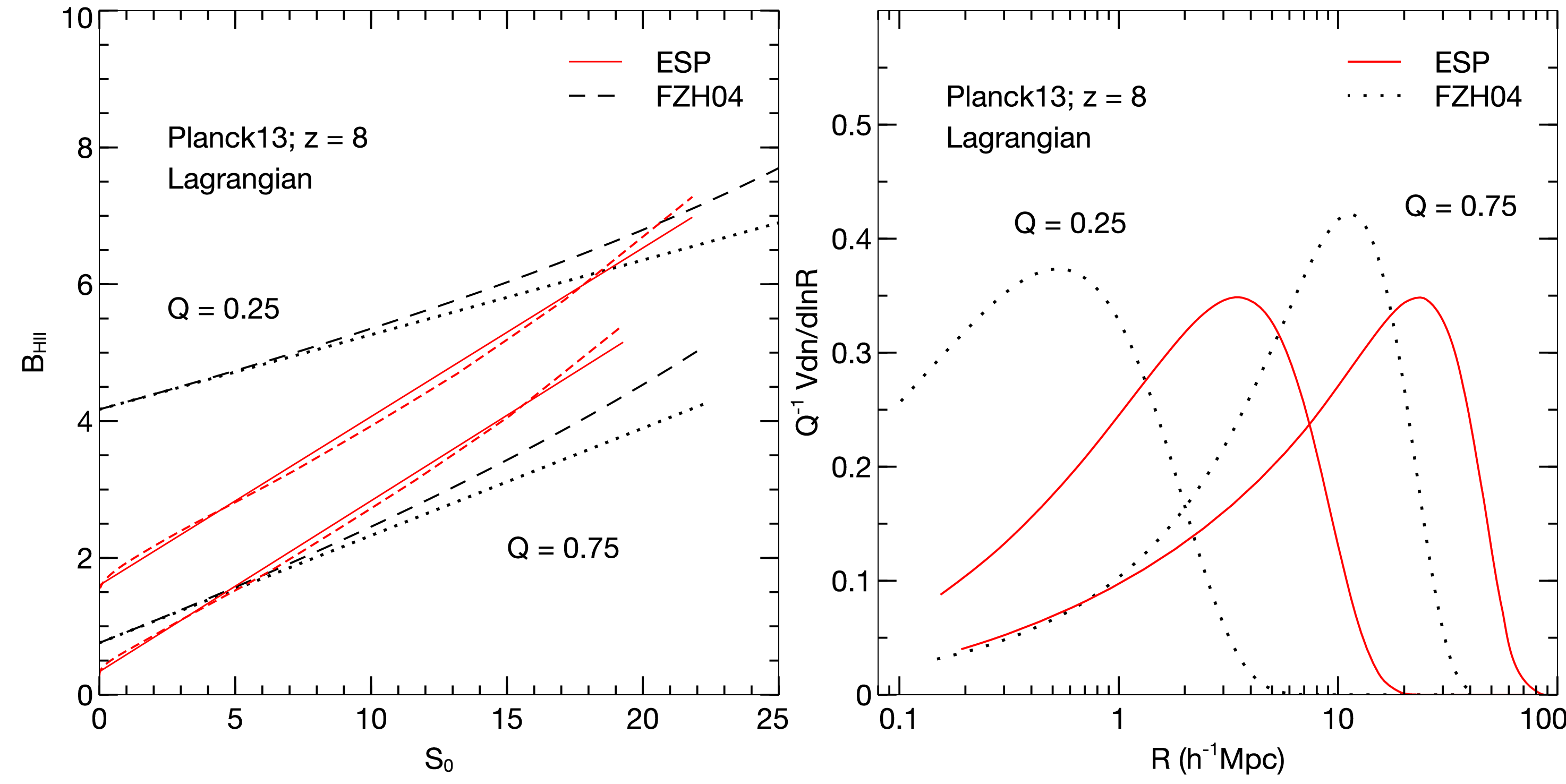}
\caption{Comparison of the FZH04 \citep{2004ApJ...613....1F} and ESP \citep{2014MNRAS.442.1470P} calculations at redshift $z = 8$ for the Planck13 cosmology at the same value of global ionized Lagrangian volume fraction. Left-hand panel: ionization barriers. Right-hand panel: normalized Lagrangian bubble size distributions corresponding to the linear barriers of the left-hand panel. See \citet{2014MNRAS.442.1470P} for details.}
\label{fig:bubbledist}
\end{figure}

These simple models of ionized bubbles do not account for the overlap and thus are valid only when the ionized fraction is very small. It is possible to account for such overlaps using the excursion set approach as was proposed by \citet{2004ApJ...613....1F}, which we refer to as FZH04. In this approach the condition for a spherical region of radius $R$ having a density contrast $\delta$ (linearly extrapolated to present epoch) to be fully ionized $\langle Q_{\rm HII} \rangle_{\delta, R} \geq 1$ is given by a condition on the collapsed fraction in the region 
\begin{equation}
\langle f_{\rm coll} \rangle_{\delta, R} \geq \zeta^{-1}.
\label{eq:fcoll_zetainv}
\end{equation}
The condition for a region to be ``self-ionized'' can be converted into a condition in terms of the density contrast $\delta$, and then problem reduces to the one for a barrier crossing. An improvement to the above model was proposed by \citet{2014MNRAS.442.1470P} by accounting for the fact that haloes would preferentially form near the density peaks \citep{2012MNRAS.423L.102M,2012MNRAS.426.2789P}, known as the excursion sets peak (ESP) model. Fig \ref{fig:bubbledist} shows the bubble size distribution for the two models FZH04 and ESP for fixed values of the ionized fraction $Q_{\rm HII}$. The main effect of the ESP model is to predict bubbles of relatively larger sizes, a fact which was confirmed while calculating the bubble distribution from semi-numerical calculations. This implies that the power spectrum predicted from the ESP model would be very different from the earlier calculations, a fact which can play an important role in interpreting the 21~cm data.

\begin{figure}
\centering
\psfrag{C2RAY}[r][r][1][0]{{\small C$^2$-RAY}} 
\psfrag{Sem-Num e=0.0}[r][r][1][0]{{\small Sem-Num}}
\psfrag{21cmFASTL}[r][r][1][0]{{\small CPS+GS}} 
\psfrag{z=10.290, xh1=0.898}[c][c][1][0]{{$x_{\rm HI}=0.90$}} 
\psfrag{z=9.164, xh1=0.557}[c][c][1][0]{{$x_{\rm HI}=0.56$}} 
\psfrag{z=8.892, xh1=0.375}[c][c][1][0]{{$x_{\rm HI}=0.38$}} 
\psfrag{z=8.636, xh1=0.147}[c][c][1][0]{{$x_{\rm HI}=0.15$}} 
\psfrag{k3 Ph1(k)}[c][c][1][0]{$k^3 P^s_0(k)/(2\pi^2)\,({\rm mK^2})$}
\psfrag{k (Mpc)}[c][c][1][0]{$k\, ({\rm Mpc}^{-1})$}
\includegraphics[width=1\textwidth,angle=0]{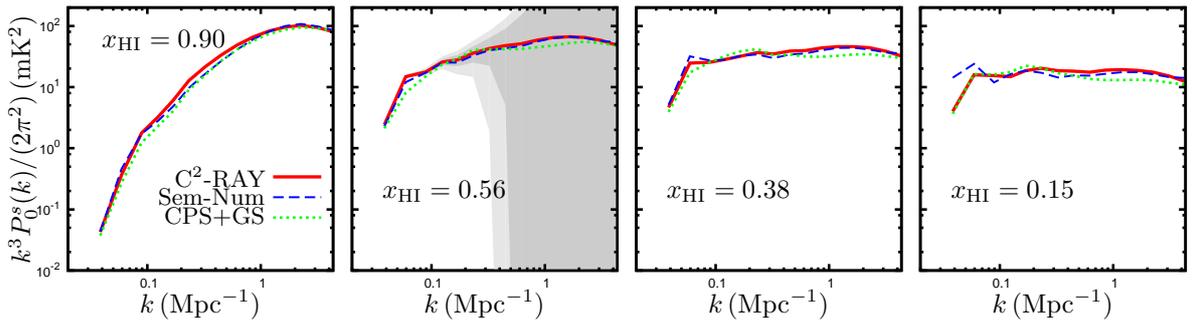}

\caption{The spherically averaged power spectrum of the redshift space  21-cm signal for the radiative transfer simulation C$^2$-RAY and two different semi-numerical models. The shaded regions in light and dark gray represent the
  uncertainty due LOFAR-like system noise at $150$ MHz for $1000$ and
  $2000$ hr of observation respectively. See \citet{2014MNRAS.443.2843M} for more details.}
\label{fig:pk_h1_spec}
\end{figure}

\subsection{Semi-numerical models}

Although the excursion-set based analytical calculations provide a reasonable description of the growth of HII regions during reionization, it is difficult to incorporate various complexities into the analytical framework e.g., the non-sphericity of bubble during overlap, the effects of line of sight peculiar velocities, quantifying the cosmic variance. It is often useful to have realisations of the ionization field so that one can construct realistic radio maps and other quantities relevant to the observations. One can use the radiative transfer simulations to generate such maps, however, they are often not suited for exploring the parameter space.

A compromise has been proposed where it is possible to generate HI maps without running the full simulation. These semi-numerical methods are based on the excursion set formalism discussed in the previous section and can be used for generating reasonably large volumes in a small amount of time \citep{2007ApJ...669..663M,2008MNRAS.386.1683G}. One of these methods is based on generating the DM density field using the perturbation theory, calculating the collapsed fraction within each grid cell (of size $R$ and density contrast $\delta$) using the analytical expression for the conditional mass function and then generating the ionization field using the excursion set formalism \citep{2011MNRAS.411..955M}. A slightly different approach is to run a full dark matter only $N$-body simulation and identify the haloes using a suitable group-finder algorithm \citep{2007ApJ...654...12Z,2009MNRAS.394..960C}. The method for generating the ionization field remains the same, i.e., using the excursion set formalism.

To understand how these semi-numerical models work, let us assume that we have a realisation of the density field along with the collapsed fraction at each grid point of the simulation box. The basic idea is to compute the spherically averaged collapsed fraction $\langle f_{\rm coll} \rangle_{R}$ for each grid point in the box for a wide range of $R$ values. If the self-ionization condition (\ref{eq:fcoll_zetainv}) is satisfied for any $R$, then the grid point is flagged as ionized. Points which fail to satisfy the condition are assigned a neutral fraction $\zeta \langle f_{\rm coll} \rangle_{R_{\rm cell}}$, where $R_{\rm cell}$ is the size of the grid cell typically set by the resolution of the map. The procedure is repeated for all points in the simulation volume. The method thus provides a realisation of the ionization field for a given value of $\zeta$. 

One important physical process which is not taken into account in the above discussion is the recombination. A significant fraction of the ionizing photons may be lost in ionizing the recombined hydrogen atoms, which may lead to very different HI topology. In the simplest case where recombinations are taken to be spatially homogeneous, the effect can be absorbed in the definition of $\zeta$. In other words, one can rewrite the ionization condition as
\begin{equation}
\langle f_{\rm coll} \rangle_{R} \geq \zeta^{-1} (1 + \bar{N}_{\rm rec}) \equiv \zeta_{\rm eff}^{-1},
\label{eq:fcoll_zetainv_nrec}
\end{equation}
where $\bar{N}_{\rm rec}$ is the average number of recombinations per hydrogen atom. Since $\bar{N}_{\rm rec}$ is assumed to be independent of the spatial point under consideration, the above equation is simply the earlier ionization condition (\ref{eq:fcoll_zetainv}) with the redefinition $\zeta \to \zeta_{\rm eff}$.

In reality, however, the recombinations are not homogeneous. In fact, the high density regions will tend to recombine faster and will be able to ``shield'' themselves from the ionizing radiation. These regions would act as ``sinks'' of ionizing photons. One way of implementing the effect of such sinks into the semi-numerical models is by including an additional condition for ionization \citep{2009MNRAS.394..960C}
\begin{equation}
\langle n_{\gamma} \rangle_{R} \geq n_H~\frac{\epsilon t_H}{t_{\rm rec}}~\frac{L}{R},
\label{eq:fcoll_zetainv_ss}
\end{equation}
where $\langle n_{\gamma} \rangle_{R}$ is the number density of ionizing photons averaged over the region of radius $R$, $t_{\rm rec}$ is the recombination time-scale, $\epsilon t_H$ is the time-scale over which the recombination term has significant contribution with $t_H$ being the Hubble time and $L$ is the comoving size of the absorbing region, usually taken to be the local Jeans scale. The effect of these recombinations is that the reionization becomes outside-in once a substantial fraction of the IGM is ionized (in contrast to models without sinks where the process is always inside-out). The presence of sinks predict smaller amplitude of fluctuations at scales $k \sim 0.1$ Mpc$^{-1}$ accessible to first generation of 21~cm experiments.

\begin{figure}
\psfrag{SNR}[c][c][1][0]{{\Large SNR}}
\psfrag{sqrt Nk}[c][c][1][0]{{\Large $\sqrt{N_k}$}}
\psfrag{k (Mpc-1)}[c][c][1][0]{\Large $k\, \, ({\rm Mpc}^{-1})$}
\psfrag{xhi=0.8}[c][c][1][0]{{\large $\bar{x}_{\rm HI}=0.8$\, \, \,}}
\centering
\includegraphics[width=.55\textwidth]{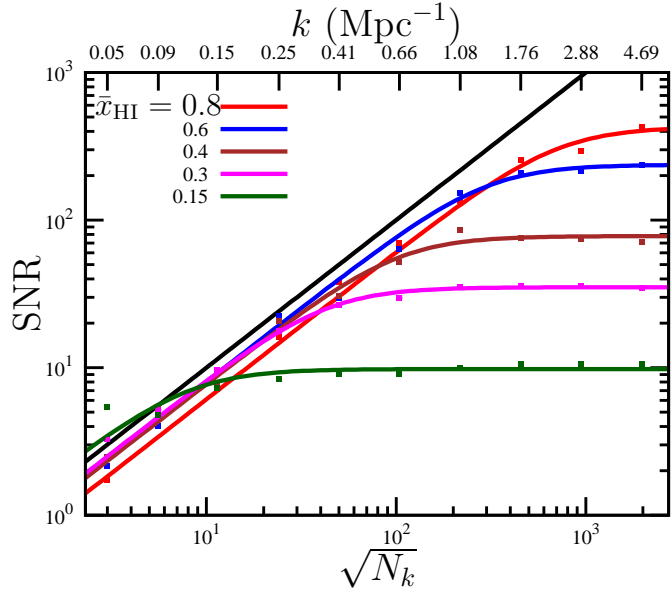}
\caption{The SNR as a function of $\sqrt{N_k}$. The 
$45^{\circ}$ black line shows the SNR expected for a Gaussian 
random field. For the $\bar{x}_{\rm HI}$ values mentioned in 
the figure, the data points (squares) show the simulated SNR 
and the solid lines show the fit given by Equation 1 of (Ref. 
Mondal et. al 2015). The $k$ value corresponding to each $k$ 
bin is shown in the top $x$ axis.}
\label{fig:snr}
\end{figure}
\begin{figure}
\psfrag{k}[c][c][1][0]{\Large $k\, \, ({\rm Mpc}^{-1})$}
\psfrag{cov}[c][c][1][0]{\Large $r_{ij}$}
\centering
\includegraphics[width=.45\textwidth, angle=-90]{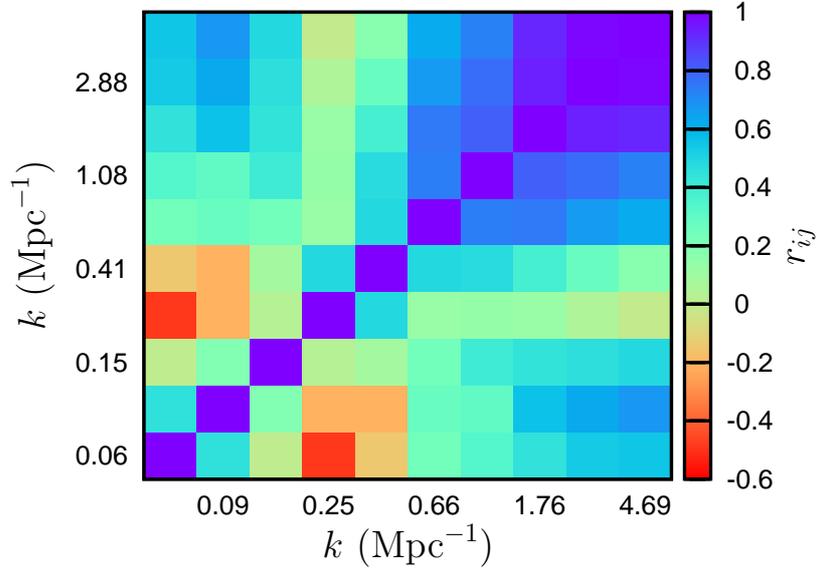}
\caption{The correlation coefficient $r_{ij} = C_{ij}/\sqrt{C_{ii}C_{jj}}$, where $C_{ij}$ is the EoR 21-cm power spectrum error covariance.}
\label{fig:cov}
\end{figure}

This model of \citet{2009MNRAS.394..960C} was used by \citet{2013MNRAS.434.1978M} to study the effect of peculiar velocities on the signal. It was proposed that the magnitude and nature of the ratio between the quadrupole and monopole moments of the power spectrum ($P_2^s /P_0^s$) can be a possible probe for the epoch of reionization. The same semi-numerical models have also been used for showing that the 21 cm anisotropy is best measured by the quadrupole moment of the power spectrum, i.e., it evolves predictably as a function of $Q_{\rm HII}$ \citep{2016MNRAS.456.2080M}.

Since the semi-numerical simulations are quite fast in terms of their computing time, they are suitable for studying various statistical properties of the signal which may require large number of realisations. One such application is to study the effects of non-Gaussianity of the ionization field on error predictions for the power spectrum \citep{2015MNRAS.449L..41M,2016MNRAS.456.1936M}. Unlike the Gaussian case where SNR $\propto N_k^{1/2}$, $N_k$ being the number Fourier modes in a given $k$-bin, the SNR has upper limit which cannot be exceeded by increasing $N_k$ \citep{2015MNRAS.449L..41M} which can be seen in Fig \ref{fig:snr}. This could have severe implications for estimating the sensitivities for the future 21~cm experiments. In fact, one can see from Fig \ref{fig:cov} that the error covariance is not diagonal any more and the coupling between different $k$-modes cannot be neglected for error estimations \citep{2016MNRAS.456.1936M}.

\subsection{Numerical simulations}

\begin{figure}
\centering
\includegraphics[width=0.9\textwidth]{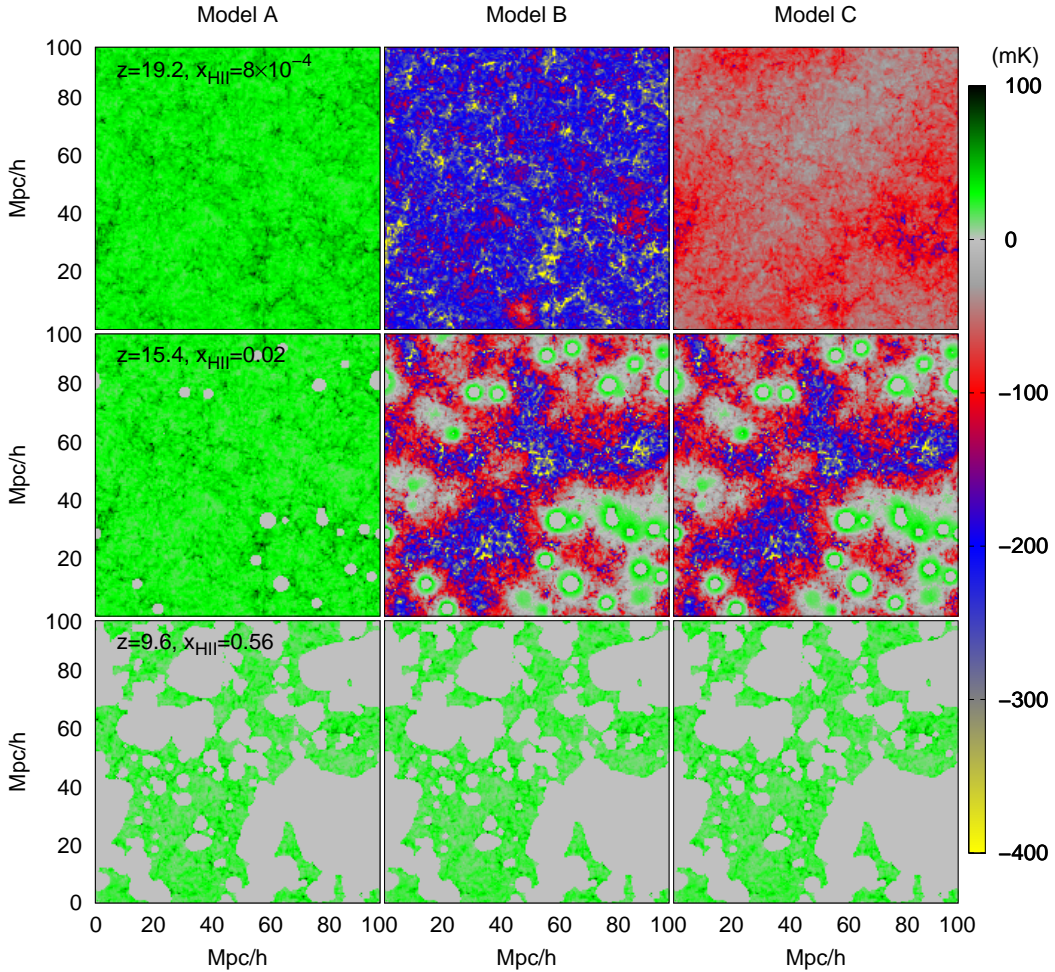}
\caption{The brightness temperature maps for three different redshifts  for three different models A, B and C. Model A assumes the IGM to be Ly$\alpha$ coupled and highly heated ($T_S \gg T_{\rm CMB}$). Model B assumes the IGM to be strongly Ly$\alpha$ coupled but self-consistently heated by X-rays, while model C considers Ly$\alpha$ coupling and X-ray heating self-consistently. See \citet{2015MNRAS.447.1806G} for details.}
\label{fig:21cm_maps}
\end{figure}

\begin{figure}
\centering
\includegraphics[width=0.55\textwidth]{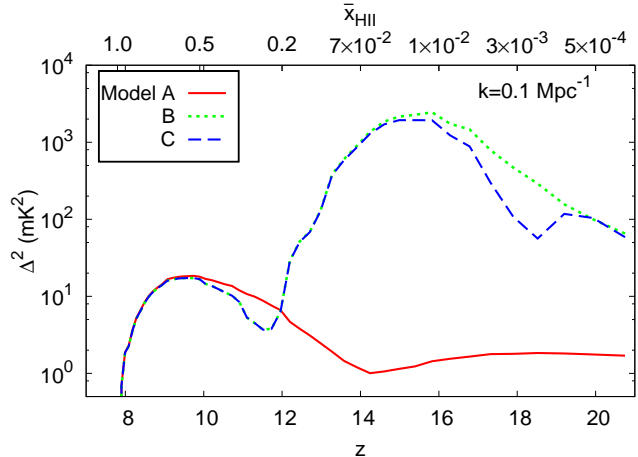}
\caption{Evolution of 21~cm power spectrum at scale $k = 0.1$ Mpc $^{-1}$ for models A (solid red), B (dotted green) and C (dashed blue), respectively. See \citet{2015MNRAS.447.1806G} for details.}
\label{fig:psz}
\end{figure}

The full complexities of the radiative transfer through the clumpy IGM can only be taken into account through the numerical solution of the radiative equation (\ref{eq:radtrans_local}). However, radiative transfer simulations are still computationally extremely challenging and hence the equation is usually solved under reasonable approximations. One such algorithm has been implemented in the radiative transfer code ``Conservative Causal Ray-tracing method'' (C$^2$-RAY) which works by tracing rays from all sources and iteratively solving for the time evolution of the ionized hydrogen fraction \citep{2006NewA...11..374M,2006MNRAS.369.1625I}. It turns out that the ionization fields generated by the C$^2$-RAY have properties which are quite similar to the semi-numerical calculations described in the previous section \citep{2014MNRAS.443.2843M}. Fig \ref{fig:pk_h1_spec} shows the 21~cm power spectrum obtained from the C$^2$-RAY compared with various semi-numerical schemes. The match turns out to be quite reasonable thus showing that one can possibly use the fast semi-numerical models to generate the 21~cm maps. 

An alternate, relatively faster, method is to use the density field and haloes from a dark matter only $N$-body simulation, and post-process with a spherically symmetric one-dimensional radiative transfer algorithm \citep{2008MNRAS.384.1080T,2009MNRAS.393...32T}. The main idea is to generate spherically symmetric 21~cm patterns around individual sources (galaxies) and then account for the overlap of such regions suitably. This method has been implemented by \citet{2015MNRAS.447.1806G} whose code not only accounts for ionization of hydrogen and helium, but also the effects of X-ray heating and Ly$\alpha$ radiation. Inclusion of these effects allows one to estimate the spin temperature at different points self-consistently and hence study the effect of fluctuations in $T_S$. Some sample outputs from the code are shown in Fig \ref{fig:21cm_maps} for three different models of treating the spin temperature fluctuations. The corresponding evolution of the 21~cm signal fluctuations at a scale $k \sim 0.1$ Mpc$^{-1}$ is shown in Fig \ref{fig:psz}. If we concentrate on model C where the heating and the Ly$\alpha$ coupling are calculated self-consistently, we find that the evolution shows three distinct ``peaks''. They arise from different physical processes, i.e., the one at the lowest redshift $z \sim 9$ arises from fluctuations in $x_{\rm HI}$, the second peak at $z \sim 15$ corresponds to fluctuations in the X-ray heating and the one at $z \sim 20$ is from the Ly$\alpha$ coupling fluctuations. Measuring these peak-like features could be quite important in understanding the physical processes at early times. It should however be kept in mind that various line of sight effects may affect the amplitude of these peaks and hence care should be maintained while interpreting the results \citep{2015MNRAS.447.1806G,2015MNRAS.453.3143G}.

\begin{figure}
\centering
\includegraphics[width=0.9\textwidth]{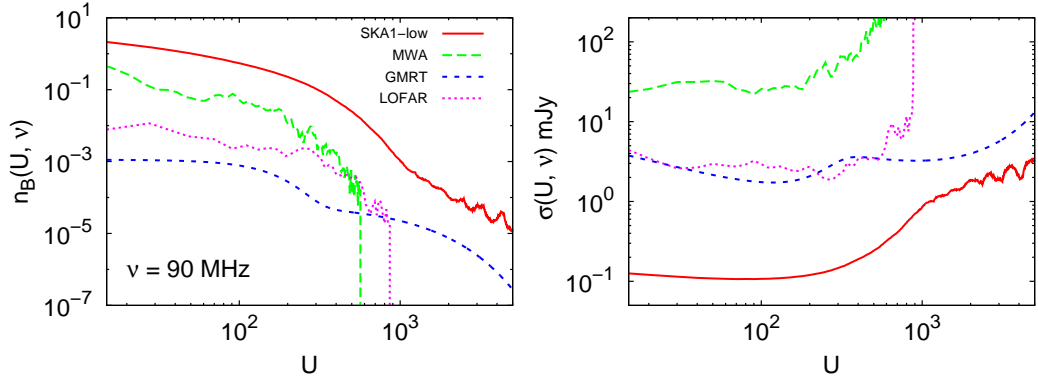}
\caption{The baseline distributions (left panel) and the corresponding rms noise in visibilities (right panel) for the SKA1-low, MWA, GMRT and LOFAR at frequency 90 MHz. $n_{\rm B}(U,\nu)$ denote the number of antenna pairs having same baseline $U$ at frequency $\nu$. The rms is computed for an observation time of  1000 hours and frequency resolution of 50 kHz. See \citet{2015arXiv151107448G} for details.}
\label{fig:ska_baseline}
\end{figure}

\section{Future outlook}

There has been tremendous progress in modelling the physics of cosmic dawn and reionization in recent times. It is now possible to construct models that are consistent with all available observations related to reionization \citep{2015MNRAS.454L..76M}. The prediction of the 21~cm signal is possibly a greater challenge and there are various approaches in modelling this.

The construction of the SKA-low phase 1 will allow us to probe the 21~cm cosmological signal with unprecedented sensitivity. Fig \ref{fig:ska_baseline} compares the expected sensitivity of SKA1-low with some of the existing facilities. As one can see, the noise variance for the SKA1-low will be almost a order of magnitude better than any of the existing ones, thus allowing us to probe the high-$z$ universe in much more detail.

Given this advancement, we require equally advanced techniques in modelling the signal so as to interpret the data accurately. Some of the directions in which the models discussed in the article can be improved are as follows:

\begin{itemize}

\item Since the physical processes at high redshifts are uncertain, they need to be parametrized by a number of free parameters (and functions). The analytical calculations could be quite helpful in probing the space of these unknown parameters as they are fast. The challenge then would be to improve the analytical models and make them as accurate as possible, most likely by comparing with semi-numerical and radiative transfer simulations. There also remain some conceptual issues with these excursion set based approaches, e.g., violation of photon conservation \citep{2015arXiv151201345P}.

\item One important physical process which is still not accounted for satisfactorily is the effect of high density regions or sinks of ionizing photons. Appropriate self-consistent and accurate treatments of these regions are required for analytical, semi-numerical and radiative transfer simulations. This could turn out to be a major challenge for reionization models in the near future \citep{2015MNRAS.446..566M,2015MNRAS.452..261C,2016MNRAS.458..135S}.

\item In addition to the 21~cm experiments, there would be instruments at other wave bands which are likely to come up in the near future probing the high-$z$ universe, e.g., JWST, TMT and so on. The models of galaxy formation as well as HI distribution should be sufficiently broad so as to explain all these observations simultaneously. This would be very important for understanding the EoR.

\item The sensitivities of the SKA1-low would not only probe the power spectrum of HI fluctuations, but would also allow us to image the HI distribution. It is thus important to explore the new information one can obtain through these images.

\item With the improvement in the noise properties of the instrument, it becomes important to devise advanced statistical estimators which can be used for obtaining the relevant quantities of interest. Thus a better modelling of the system noise coupled with improved reionization models is required in the near future.

\end{itemize}

\bibliographystyle{astron}
\bibliography{eor_models}

\end{document}